\documentclass[twocolumn]{aastex63}

\usepackage{tabularx}
\usepackage{url}
\usepackage{graphicx}
\usepackage[T1]{fontenc}
\usepackage{ae,aecompl}
\usepackage{booktabs}
\usepackage{natbib}
\usepackage{multirow}
\usepackage{multirow}
\usepackage{appendix}
\usepackage{hyperref}

\DeclareGraphicsExtensions{.pdf,.png,.jpg,.mps,.eps,.ps}
\usepackage{amsmath}
\usepackage{blindtext}
\usepackage{longtable}
\usepackage{tabu}
\usepackage{lineno}
\usepackage{chngcntr}
\usepackage[utf8]{inputenc}
\usepackage{orcidlink}
\usepackage{rotating} 

\usepackage{listings}
\usepackage{color}

\definecolor{mygray}{gray}{0.95}

\def\aap{A\&A} \def\apjl{ApJ}  
 \def\apjs{ApJS}

 \submitjournal{ApJ}

\shorttitle{Time Resolved Spectroscopy}

\begin{document}

\title{The response of warm absorbers to the variations in the ionizing continuum in the active galaxy NGC~4051}

\author[0000-0002-9163-8653]{Dev R. Sadaula} 
\affiliation{Astrophysics Science Division, NASA Goddard Space Flight Center, Greenbelt, MD 20771, USA.}
\affiliation{Center for Space Science and Technology, University of Maryland Baltimore County, 1000 Hilltop Circle, Baltimore, MD 21250, USA.}

\affiliation{Center for Research and Exploration in Space Science and Technology, NASA/GSFC, Greenbelt, Maryland 20771, USA}

\author[0000-0002-5779-6906]{Timothy R. Kallman} 
\affiliation{Astrophysics Science Division, NASA Goddard Space Flight Center, Greenbelt, MD 20771, USA.}

\author[0000-0003-2714-0487]{Sibasish Laha} 
\affiliation{Astrophysics Science Division, NASA Goddard Space Flight Center, Greenbelt, MD 20771, USA.}
\affiliation{Center for Space Science and Technology, University of Maryland Baltimore County, 1000 Hilltop Circle, Baltimore, MD 21250, USA.}
\affiliation{Center for Research and Exploration in Space Science and Technology, NASA/GSFC, Greenbelt, Maryland 20771, USA}

\correspondingauthor{Dev R. Sadaula}
\email{dev.r.sadaula@nasa.gov}

\begin{abstract}
We present a time-resolved X-ray spectral analysis of the warm absorbers in the Seyfert galaxy NGC 4051, which has an active galactic nucleus (AGN), using observations from the Neutron Star Interior Composition Explorer (NICER). Despite NICER’s moderate spectral resolution, its high-cadence monitoring allows us to probe the response of the ionized outflows, also known as warm absorbers, on timescales of $\sim$5500~s. We detect two distinct components of ionized absorbers in this source. The variability in the ionization parameter of the low-ionization warm absorber component, which tracks changes in the ionizing flux with no measurable time lag. This rapid response implies photoionization equilibrium and places a lower limit on the electron density of $\gtrsim 9\times10^6$~cm$^{-3}$ based on the most abundant ionic species, O VII. The absorber is located within $\sim 0.02$~pc of the central source, consistent with an origin in the inner regions of the active nucleus. In contrast, the high-ionization absorber remains consistently under-ionized relative to equilibrium predictions. This suggests that it may be the collisional plasma, which was also detected in this source in the previous studies. These results demonstrate that time-resolved spectroscopy, even with moderate-resolution instruments, can provide valuable constraints on the density and location of warm absorbers in AGN. As a potential candidate source of AGN feedback, the study of these ionized outflows is crucial in understanding AGN-host galaxy interactions.
\end{abstract}


\keywords{Galaxy, Seyfert, X-rays, AGN, \texttt{{\sc {\sc warmabs}}}, Black hole, Photoionization}

\vspace{0.5cm}

\nolinenumbers

\section{Introduction}

Active Galactic Nuclei (AGN) are among the most luminous and energetic objects in the universe, powered by the accretion of matter onto supermassive black holes. A significant fraction of AGNs exhibit ionized outflows, known as warm absorbers (WAs), which are detected through absorption features by partially ionized gas in their X-ray spectra \citep{roz05,Behar2003,cre03,krolik95,hal84}. Typical WA outflow speeds are 500 -- 1000 km s$^{-1}$. Some observations reveal ultrafast outflows with much larger speeds (0.1 - 0.3c) \citep{cha02,pau103,pau203,tom10}.  These outflows are surprising, given the fact that we expect primarily an inflow toward the black hole to fuel the AGN. Studies show that these outflows have the potential to influence galaxy evolution by regulating star formation and redistributing energy throughout the host galaxy, called the AGN feedback mechanism \citep{elv00,cre03,blu05,rosito2021}.

The WA gas is heated and ionized by the strong UV and X-ray continuum radiation originating from close to the central black hole. If it is assumed that the gas is in  ionization and thermal equilibrium, then WAs  are characterized by their ionization parameter, $\xi=L/n r^2$, where $L$ is the ionizing luminosity of the source and $n$ is the density of WA \citep{tarter69}. We can directly infer the WA parameters from observed spectra by comparing models for the ionized absorbers.  This yields the WA parameter values:  log$(\xi) \simeq -0.5 $ - $  3.0 $; column density, N$_\text{H}\simeq (10^{20}-10^{22})$ cm$^{-2}$ and outflow speed, $v_{\text{out}}\simeq (500-2000)$ km s$^{-1}$ \citep{kas02, sibwax_2014}. Some recent studies have shown much higher speeds \citep{Ogorzalek2022}. These values are used to calculate the mass flux in the outflow if the geometry of the outflow is assumed.  This in turn suggests that WA outflows can carry significant amounts of mass and energy, potentially affecting the host galaxy on both small and large scale \citep{silk98,dimatteo2005,fab12,cro06,cat09}.
However, the origin and acceleration of WAs are unclear.  Possible driving mechanisms include radiatively driven outflow, magnetically driven outflow,  thermal evaporation from a cold accretion flow, or a combination of all of the above. 

Study of the WA is usually performed using an assumption of equilibrium approximation. However, extreme variability of the ionizing luminosity near the black hole forces the gas to go out of equilibrium.  Previous studies \citep{rog22, sadaula2023, sadaula2024, gar13, kro07, nic99} have shown the absorbers are out of equilibrium in some of the sources. The departure of the equilibrium depends on the time scale of variability of the ionizing source, and the characteristic timescales affecting ionization, recombination, and other atomic processes depend on fundamental atomic constants, together with the density and the ionizing flux. Time-resolved spectral analysis provides an additional powerful technique to study  WAs. We can probe the density of the absorber independently using this technique.

Departures from ionization equilibrium can be detected by comparing observations with time-dependent models. Models assuming ionization equilibrium predict specific ratios of the strengths of absorption lines in the spectrum; observed spectra that do not match these predictions are likely out of equilibrium. Fits of such spectra to models incorporating time-dependent effects can be used to infer key physical quantities affecting time dependence, namely the gas density and the radiation flux incident on the WA. Development of time-dependent photoionization models has been carried out by \citet{nic99, kro07, krolik95, Kaastra2012, rog22, gar13, sadaula2023, sadaula2024}. The implementation of the time-dependent photoionization method and applications of these models to observed data have been later conducted by \citet{Luminari2023, lum24, gu23, juranova2022}.

In our previous work \citep{sadaula2023, sadaula2024} and work by others \citep{rog22,li23,gu23,nic99,kro07}, we have shown the variability of the central ionizing source could potentially be a useful probe for the density diagnostic of the ionizing outflow. All  this work has explored a similar method to calculate time-dependent photoionization and recombination effects. That is, the time-dependent ionization is calculated  due to variations in the ionizing luminosity.  Then model spectra vs. time are produced for different densities and fluxes. These models are then fit to observations in order to search for the signatures of time-dependent photoionization. Such signatures, if found, can be used to constrain density and flux. This procedure is challenging owing to the fact that it requires looking at the time variability of several, or many, spectral features (i.e., lines or edges) simultaneously. In what follows, we will refer to this method as `time-dependent photoionization model fitting.'
 
Another way to detect time-dependent ionization is to make use of a simplified heuristic model of time-dependent photoionization. That is, if the gas is initially in photoionization equilibrium with the observed radiation field, and then if the radiation field suddenly increases or decreases to a different constant value, the ionization balance in the gas will respond.  The response will resemble a changing ionization parameter, with the state of the gas accurately characterized by the initial ionization parameter prior to the change in flux and by the final ionization parameter long after the change in flux.  The transition between the initial and final states will take more time than the (sudden) change in the illuminating flux from the initial to the final values. We call this response time the equilibrium time, $t_{eq}$ and it depends on an initial equilibrium photoionization timescale $t_{PI}$ and the recombination timescale, $t_{Rec}$. These definitions are adopted from the previous work by \citet{gar13, sadaula2023}. Thus, by measuring the change in the ionization parameter from a time series of spectra, we can set limits on $t_{eq}$, which is technically the time lag between the variation in the ionizing flux and the ionization state of the gas. The equilibrium time directly depends on the gas density and can be used to determine its value. This is the procedure that we adopt in this paper. 
\begin{figure*}
    \centering
    \hbox{
    \includegraphics[width=0.8\linewidth]{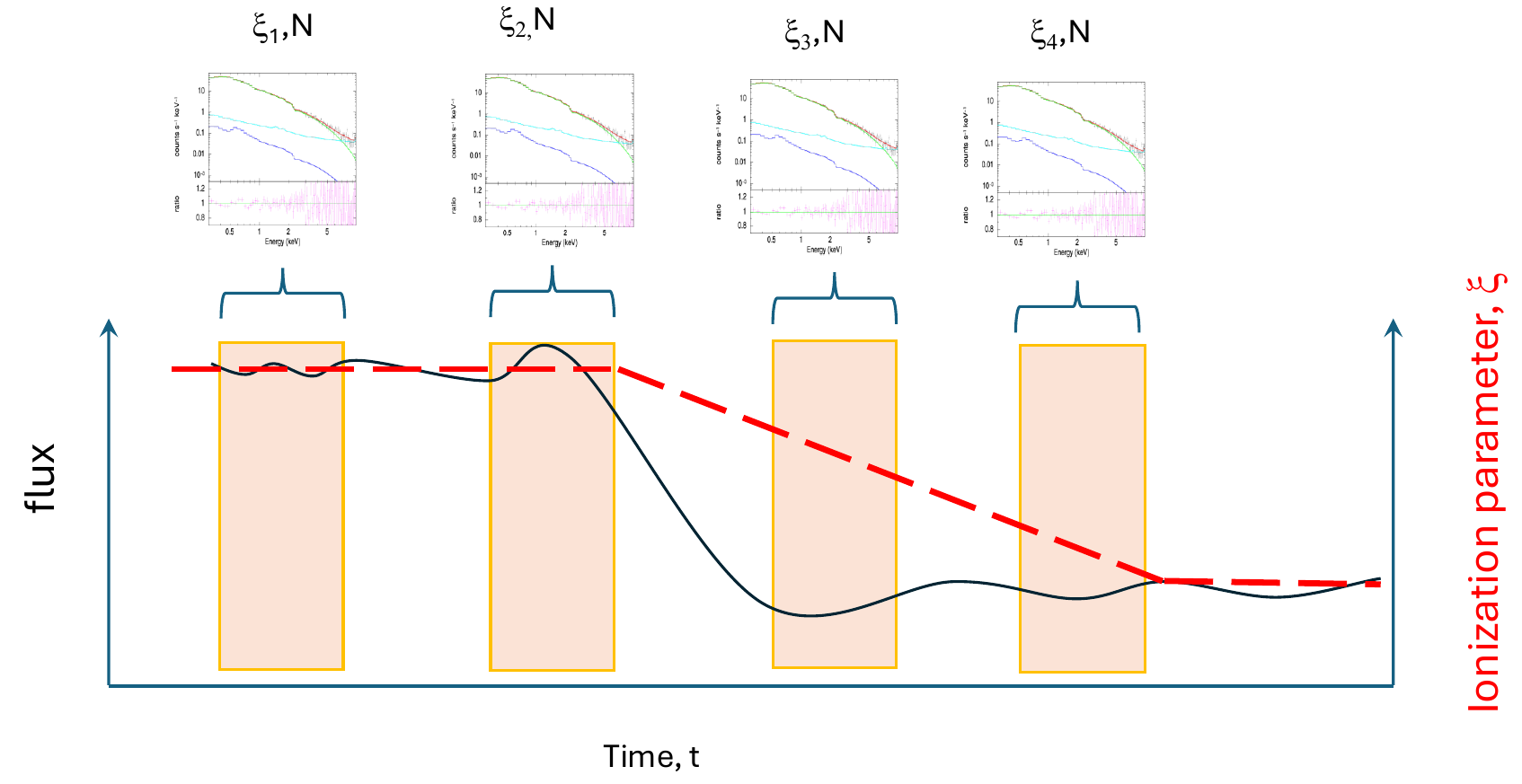}
    }
    \caption{Schematic drawing showing our procedure for studying time-dependent ionization. We divide the observation into intervals and fit for the ionization parameter vs. time (as shown by the red curve) and test for correlation between the change in the ionization parameter and the flux (shown as the black curve). We interpret the lag between the change in the continuum vs. the ionization parameter as the ionization or recombination time.  The hatched regions denote the time intervals needed in order to accumulate spectra capable of measuring the ionization parameter.}
    \label{fig:spec1}
\end{figure*}

Figure \ref{fig:spec1} illustrates schematically how our procedure works.  We create time-resolved spectra  and fit each spectrum with an equilibrium WA model. We then compare the WA ionization parameter with the continuum flux in each spectrum.  We do this by calculating the cross-correlation function, including a variable lag.  In this way we search for a lag between the ionizing continuum and the warm absorber gas.  If found, such lags most likely imply departures from ionization equilibrium.  We demonstrate, with simulations, that this test will detect time dependence even though it employs equilibrium models to model WA to provide the ionization parameters for the various time bin spectra.  

Our method has advantages compared with the "time-dependent photoionization model fitting." This method measures the conditions in the equilibrium states before and after a change in ionizing flux and then compares the time separation of the equilibrium states with the predictions of time-dependent photoionization.   Detection and characterization of the ionization balance in equilibrium models requires fewer photons than does the full characterization of time-dependent models.  Detection of time dependence in a single spectrum requires detection of two or more spectral features and then demonstrating that their ratios cannot be fitted by equilibrium models.   Detection of a warm absorber in equilibrium can be as simple as detecting one edge or line feature and showing that it is stronger than can be produced by cold absorption.   Thus, our method requires fewer counts than direct detection of time dependence in each spectrum.  Therefore, it can probe shorter timescales and is conceptually simpler.

In this paper, we use an observation from NICER \citep{keith2016} to investigate the warm absorber in NGC 4051 and its response to variations in the ionizing flux. NGC 4051, a well-studied Narrow Line Seyfert 1 galaxy, exhibits significant X-ray variability and hosts a complex warm absorber (WA) structure with at least two distinct ionization states \citep{kro07}. Previous studies \citep[e.g.,][]{sil16, Collinge2001, Crenshaw2012, Pounds2004, Ogorzalek2022} have revealed multiple absorption components with distinct ionization parameters, column densities, and velocities. Recent work \citep{Ogorzalek2022} using a Bayesian framework found some components in photoionization equilibrium and others in collisional equilibrium. Additional studies \citep[e.g.,][]{King2012, Mizumoto2017} explored the variability and launching mechanisms of these absorbers. These previous findings provide essential context for studying and interpreting WA using NICER data with moderate resolution.

The structure of this paper is as follows: Section 2 describes our warm absorber modeling approach \citep{kall01} and the construction of a realistic spectral energy density (SED). Section 3 details the NICER observation and data reduction procedures. Section 4 covers the spectral fitting and data analysis techniques. Section 5 presents the results, including warm absorber parameter estimates and cross-correlation analysis. Section 6 discusses the physical implications of these findings, while Section 7 summarizes the key conclusions, followed by Section 8, which outlines future prospects.

\section{Warm Absorber Modelling and its detection by NICER}

Modeling of WA requires an accurate treatment of the ionizing continuum of the source. For this reason, we developed a realistic SED, a model that is used to fit the ionized absorbers in the observed spectrum. We also discuss the capabilities of the NICER instrument in detecting the ionized outflow and constraining its parameters depending upon the ionization  and column of the absorbers in this section.

\subsection{Creating a Realistic SED for NGC 4051}\label{sec:sed}

\begin{figure}[h]
    \centering
    \includegraphics[width=1.\linewidth]{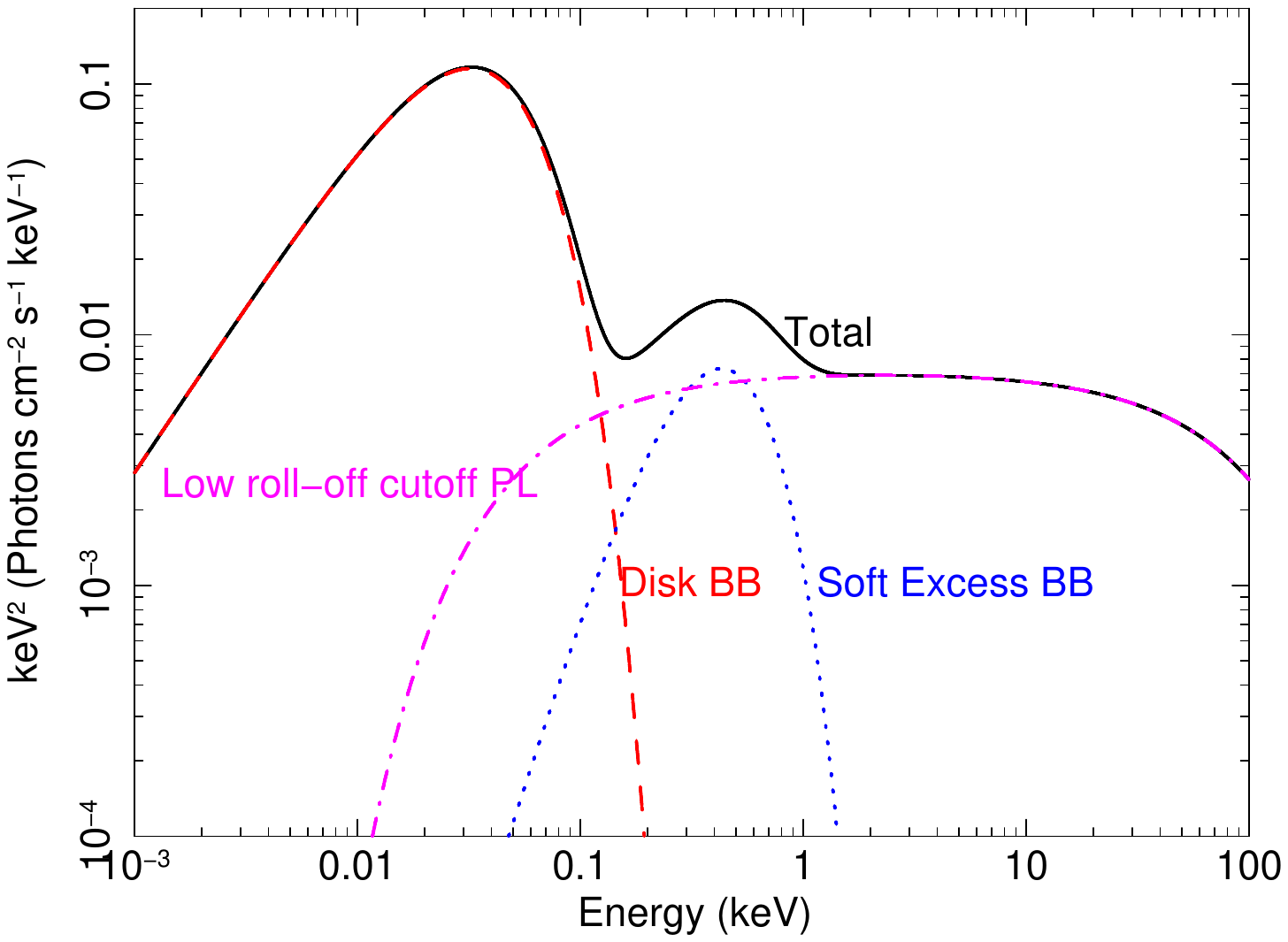} 
    \caption{Broadband SED model for NGC 4051. The red dotted curve shows the accretion disk blackbody emission with a temperature of 13.6 eV at the innermost region. The blue dotted curve represents the soft excess blackbody emission at 0.11 keV. The pink dotted curve illustrates the cutoff power-law component, with a low-energy roll-off at 50 eV, a high-energy cutoff at 100 keV, and a photon index of 2. The black solid line represents the total continuum emission for NGC 4051 as derived from this composite model.}
    \label{fig:sed}
\end{figure}

For this project, we constructed a broadband SED for NGC 4051 using a model \textsf{diskbb}+\textsf{bbody}+\textsf{cutoffpl} that combines three key components. The parameter values and normalizations for the \textsf{diskbb} component, which represents the accretion disk multi-color blackbody, were taken from \cite{nucita2010,alston2013} as a reference. The parameters and normalization for the soft excess blackbody (\textsf{bbody}) and the high-energy cutoff power-law (\textsf{cutoffpl}) were taken from \cite{kro07}. The details of the model parameters, including the temperatures, normalizations, and spectral indices, are summarized in Table \ref{table:sed_model_parameters} and visualized in Figure \ref{fig:sed}.

\begin{table}[h!]
\begin{tabular}{cccc}
\hline
\textbf{Model Comp.} & \textbf{Parameter} & \textbf{Unit} & \textbf{Value} \\
\hline
diskbb & Tin & eV & 13.6 \\
diskbb & norm & -- & 5.00E+08 \\
bbody & kT & keV & 0.14 \\
bbody & norm & -- & 1.90E-04 \\
expabs & LowECut & eV & 50 \\
cutoffpl & PhoIndex & -- & 2 \\
cutoffpl & HighECut & keV & 100 \\
cutoffpl & norm & -- & 7.20E-03 \\
\hline
\end{tabular}
\caption{Model parameters values for the broadband SED of NGC 4051.}
\label{table:sed_model_parameters}
\end{table}

\subsection{Warm Absorber Modeling}
The warm ionized gas absorption in the observed spectra is modeled using the analytic model {\sc {\sc warmabs}} \citep{wa} \footnote{available from the webpage\\ https://heasarc.gsfc.nasa.gov/docs/software/xstar/xstar.html}. This model allows XSPEC \citep{arn96} to simulate warm absorbers without requiring users to construct their own XSTAR \citep{kall01} input or output tables. This approach relies on precalculated "population files," which are standard XSTAR output files. Using these files, the \textsf{warmabs} model can directly produce the absorption in synthetic spectra, skipping the complex steps involved in solving rate equations, which is the main time-consuming component in XSTAR.

\subsection{Detection of warm absorbers with NICER  }
We explore the detectability and significance level of WA features that NICER can constrain. For this, we generated simulated spectra using the model \textsf{tbabs}$\times$\textsf{warmabs}$\times$(\textsf{bbody}+\textsf{powerlaw}) with the \textit{fakeit none} command in XSPEC using NICER response file. We adopted a power-law photon index of $\Gamma = 2$ and a blackbody temperature of 0.10 keV, consistent with the parameters used to construct the SED in Section~\ref{sec:sed}. Model normalizations were matched to the SED to ensure realistic continuum levels.

\begin{figure}[h]
    \centering
    \includegraphics[width=1.\linewidth]{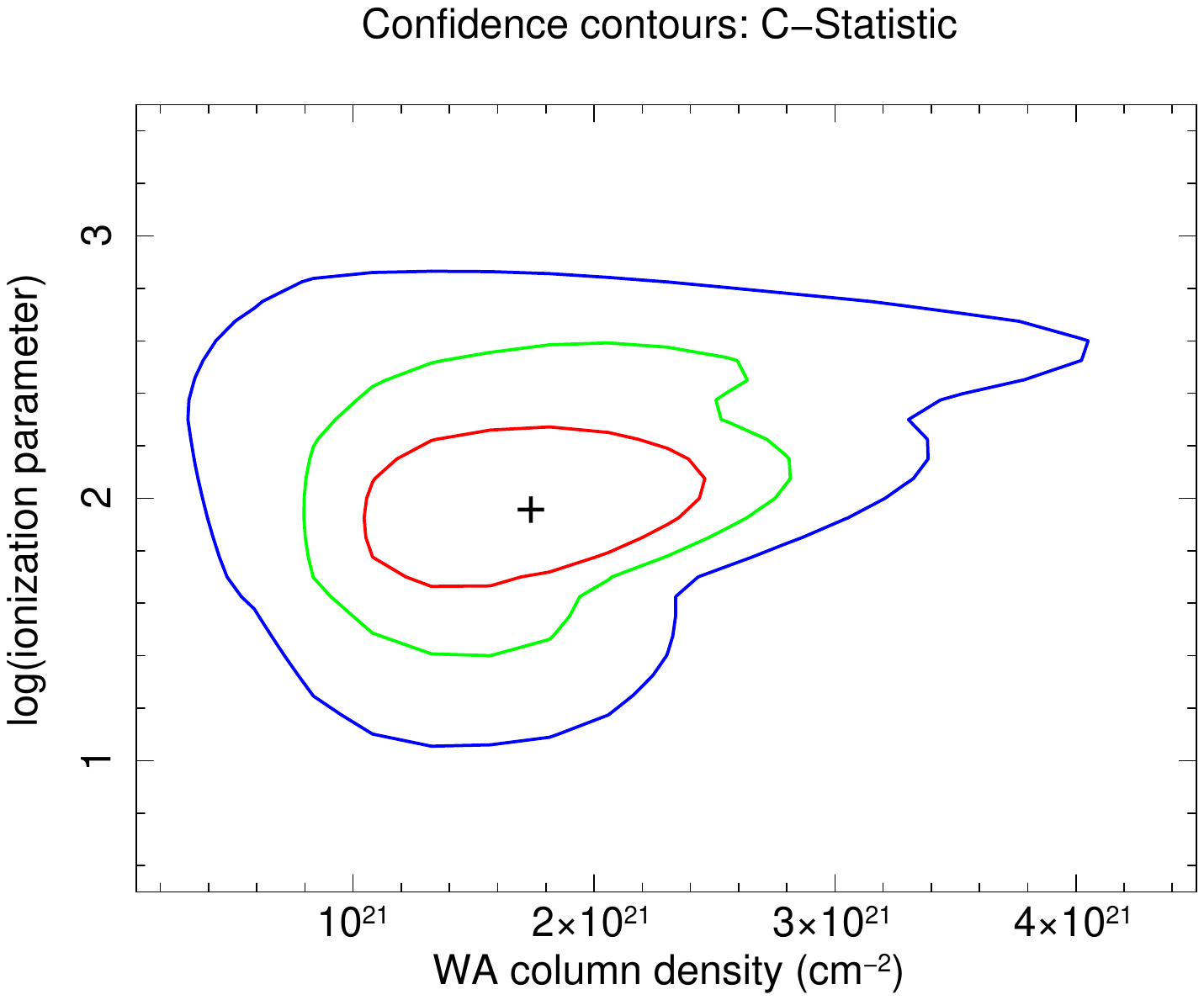} 
    \caption{Contour plots of warm absorbers parameters log(ionization parameter) and WA column density obtained by fitting the simulated spectra with column density, $2.0 \times 10^{21}$ cm$^{-2}$ and log($\xi$) =~ 2.0. Red, green, and blue correspond to the 68\%, 90\%, 99\% confidence regions.}
    \label{fig:contour_sim}
\end{figure}

For the simulation, we assumed an exposure time of 1000\,s, with the warm absorber column density fixed at $2.0 \times 10^{21}$\,cm$^{-2}$ and the ionization parameter at $\log(\xi) = 2.0$. We then fitted the simulated spectra and generated a contour plot of $\log(\xi)$ versus the warm absorber column density, as shown in Figure~\ref{fig:contour_sim}. This demonstrates that NICER can detect and constrain the warm absorber parameters with 90\% confidence for a 1000-second exposure. Therefore, any non-detection of warm absorbers in our NICER data is unlikely to be due to the instrument’s sensitivity limitations within the parameter space depicted in Figure~\ref{fig:contour_sim}.

\section{Observation, Data Reduction}

\subsection{Observation}
NICER observations of AGNs such as NGC 4051 typically consist of many individual pointings, with approximately contiguous exposure times of $\sim$ 1000s within each. These pointings are separated by longer intervals ($\sim 4500$s) due to earth occultation or the South Atlantic Anomaly (SAA).   For the purpose of this paper, we selected one observation (OBSID  1112010115), observed on 2018-01-15.
 
\begin{figure*}
    \centering
    \includegraphics[width=0.95\textwidth]{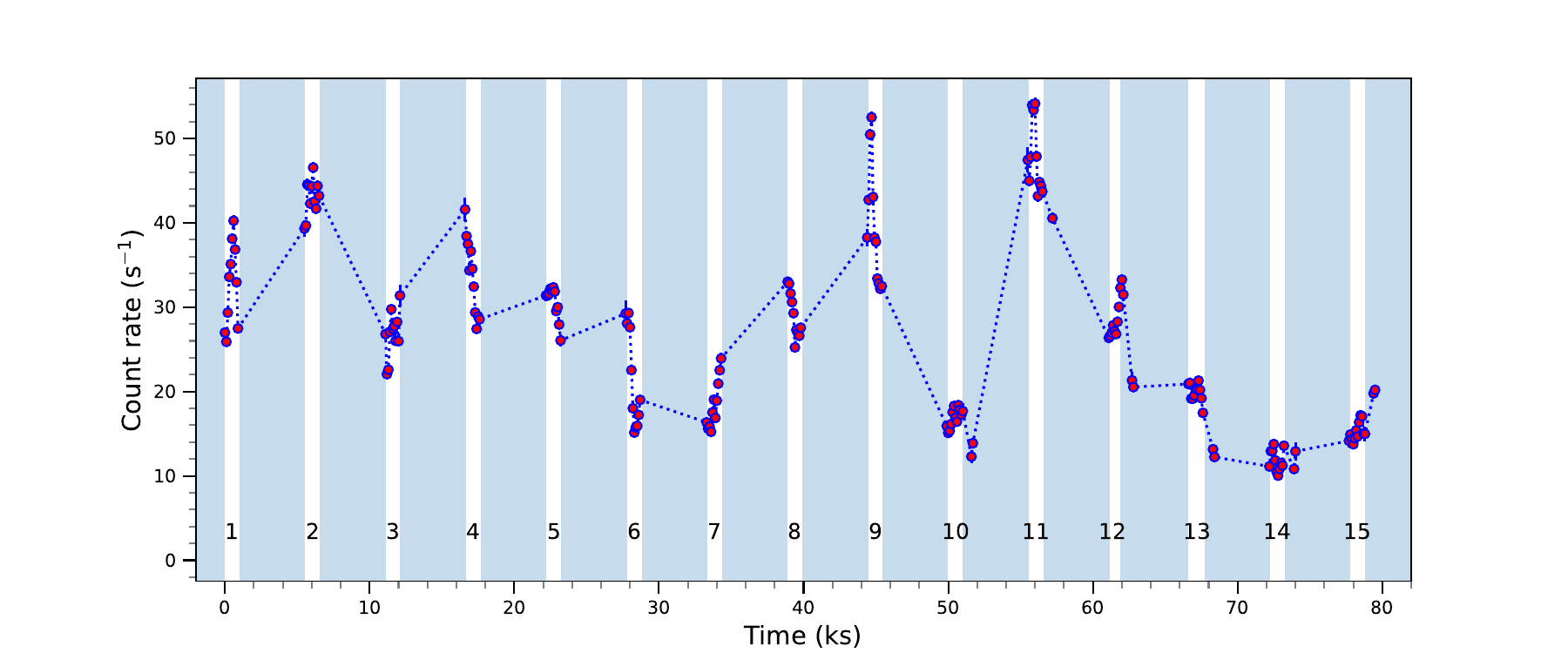}
    \caption{The light curve corresponding to observation ID 1112010115 for NGC 4051. The data gaps are shown in blue bands; each data group is given a number from 1 to 15. The light curve is binned with 100 s. Only those time points that fall in the white region are considered for the analysis.} 
    \label{fig:lc}
\end{figure*}

The details of this observation are given in table \ref{tab:gti} and visualized in figure \ref{fig:lc}. The second and third columns of table \ref{tab:gti} represent the starting and stopping times for a particular pointing, the fourth column gives the exposure for this pointing, and the last column represents the gap between two consecutive pointings. These gaps are $\sim 4500 s$. The pointings and the gaps are shown in figure \ref{fig:lc}. The telescope was pointed to the source 28 times during the whole observation. The exposure time ranged from a few seconds to 1000 seconds. For our work, we chose only those pointings with exposure greater than 700 s and discarded all other pointings because either they were very short or they were far apart from the consecutive pointings greater than 700 s. This filtering criterion gives us 15 different spectra from 15 pointings, each with an exposure of $\sim 1000 $ s or greater. These pointings are given a short name: s1, s2, s3, and so on,  as shown in Table \ref{tab:gti}.

We selected NICER for this study due to its large effective area around 1 keV, where warm absorber features are prominent, and its capability for rapid-cadence observations of variable AGN. Although Earth occultations cause regular observational gaps, NICER allows for better sampling of flux variability compared to longer uninterrupted observations with XMM-Newton. The selected observation (OBSID 1112010115) exhibited large flux variations and sufficient exposure per pointing to enable time-resolved spectral analysis. The count rates range from a minimum of $\sim 15$ counts s$^{-1}$ at a time $\sim 50.5$ ks to a maximum of $\sim 50$ counts s$^{-1}$ at a time  $\sim 56$~ks as shown in the light curve \ref{fig:lc}.  This fulfills  the criterion of significant variation of a factor of $\sim 3$ on a timescale of $\sim 5500$ s. We note that nearby observations (e.g., OBSID 1112010116) were examined but excluded because they were observed at least a day apart from each other, which may produce the spurious correlation between WA parameters.

\begin{table}[h!]
\caption{The start time, stop time, exposure, and the gap between consecutive pointings that are greater than 700 s. For example, for the first pointing (s1), also called the segment (seg), the telescope starts to observe at 127443371 s and ends at 127444364 s with a total exposure of $t=993$ s, and then there is a break in the observation for gap$=4562$~s.}
\label{tab:gti}
\centering
\begin{tabular}{ccccc}
\toprule
seg. & start time & stop time & exp. & gap \\
     & (s)        & (s)       & (s)  & (s) \\
\midrule
s1 & 127443371 & 127444364 & 993 & 4562 \\
s2 & 127448926 & 127449941 & 1015 & 4545 \\
s3 & 127454486 & 127455485 & 999 & 4563 \\
s4 & 127460048 & 127461042 & 994 & 4565 \\
s5 & 127465607 & 127466621 & 1014 & 4536 \\
s6 & 127471157 & 127472181 & 1024 & 4528 \\
s7 & 127476709 & 127477721 & 1012 & 4548 \\
s8 & 127482269 & 127483281 & 1012 & 4552 \\
s9 & 127487833 & 127488841 & 1008 & 4481 \\
s10 & 127493322 & 127494384 & 1062 & 4563 \\
s11 & 127498947 & 127499961 & 1014 & 4549 \\
s12 & 127504510 & 127505269 & 759 & 4711 \\
s13 & 127509980 & 127511081 & 1101 & 4546 \\
s14 & 127515627 & 127516641 & 1014 & 4527 \\
s15 & 127521168 & 127522182 & 1014 & --- \\
\bottomrule
\end{tabular}
\end{table}

\subsection{Data Reduction}
The data was reduced using the standard NICER pipelines, $\textit{nicerl2}$, and the cleaned event files were obtained. Then, we used $\textit{maketime}$ the HEASoft command to make a good time interval (gti) file for each observed segment greater than 700 s exposure and ran nicerl3-spect for each one of them, which generated the final spectral products ready for scientific analysis. The latest calibration file (20240206) has been used to reduce the data. The spectra in the case of NICER are binned automatically using the default method of optimal binning \citep{kaa16} when the $\textit{nicerl3-spect}$ pipeline is run.

\section{Spectral Fitting and Analysis}
Fitting time-resolved X-ray spectra requires balancing temporal resolution with statistical quality. High time resolution is desirable but limited by the need for an adequate signal-to-noise ratio (S/N) and the intrinsic time resolution of the instrument \citep[e.g.,][]{vaughan03}. These constraints are further compounded when using complex spectral models with many free parameters, which can lead to parameter degeneracies and increased statistical uncertainties \citep{arnaud96, protassov02}. To mitigate these effects, we reduce the number of free parameters in our model while fitting time-resolved spectra. Therefore, we first fit the time-averaged spectra for the entire observation and use the best-fit model to fit the time-resolved spectra, letting only ionization parameters of the WAs vary.

\subsection{Time-averaged Spectral fitting}
Prior to performing time-resolved spectral analysis, we carried out a time-averaged spectral fit using the full accumulated exposure for the observation. This fit was implemented in XSPEC using a physically motivated model that includes photoionized absorption components. We considered three spectral models of increasing complexity:

\begin{itemize}
    \item \textbf{Model 1:} \texttt{TBabs*(bbody + powerlaw)} — a baseline model that includes Galactic neutral absorption, soft excess emission modeled as a blackbody (\textsf{bbody}), and a power-law continuum.
    \item \textbf{Model 2:} \texttt{TBabs*warmabs*(bbody + powerlaw)} — a single warm absorber component (\textsf{warmabs}) to account for ionized absorption by outflowing gas in addition to Model 1.
    \item \textbf{Model 3:} \texttt{TBabs*warmabs*warmabs*(bbody + powerlaw)} — two distinct warm absorber components in addition to baseline model, Model 1.
\end{itemize}

In these models, \textsf{TBabs} accounts for Galactic neutral absorption, \textsf{warmabs} represents absorption by photoionized outflowing gas, \textsf{bbody} models the soft excess emission, and \textsf{powerlaw} describes the high-energy coronal continuum. While the \textsf{bbody} component phenomenologically captures the soft excess, its exact physical origin remains debated. Two leading interpretations are: (1) blurred reflection of the primary X-ray continuum by partially ionized, relativistically smeared inner accretion disk material \citep{Crummy2006, Walton2013, Parker2014}; and (2) emission from a warm, optically thick Comptonizing corona that upscatters disk photons into the soft X-ray band \citep{Gierlinski2004, Mehdipour2011, Petrucci2018}.

For all models, the Galactic column density was fixed at $N_{\mathrm{H}} = 1.20 \times 10^{20}$ cm$^{-2}$, consistent with the HI4PI survey \citep{hi4pi}, and data between 0.3 to 10~keV were used due to background domination outside this range. Since NICER does not have an imaging detector, the backgrounds were estimated using the "Scorpean Model." Each \textsf{warmabs} component includes a column density ($N_{\mathrm{H}}$), ionization parameter ($\xi$), and redshift $z$ (encode outflow velocity), with turbulent velocity fixed at 300 km~s$^{-1}$. In this analysis, all \textsf{warmabs} parameters, including column density, ionization parameter, and redshift, were allowed to vary freely. We report both the fit statistic (C-stat) and the test statistic ($\chi^2$) to demonstrate the goodness of the fits. However, for the remainder of this paper, we refer to the C-statistic.

The best-fit parameters for each model, along with their symmetric 90\% confidence intervals, are listed in Table~\ref{tab:time_ave_fit}. The continuum-only fit (Model 1) provides a poor description of the data, yielding a reduced chi-squared value of approximately 2. To improve the fit, we introduced a single warm absorber component (Model 2), which significantly improved both the fit and statistical quality. In Model 2, the WA redshift is measured to be $z = -0.065 \pm 0.003$, corresponding to an outflow velocity of approximately $-19{,}400$kms$^{-1}$ (where the negative sign indicates a blueshift).

To explore the possibility of multiple absorbers, we extended the model to include two warm absorber components (Model 3), which resulted in a further improvement in the fit statistics. The two WA components in Model 3 have redshifts of $z = -0.013$ and $z = -0.034$, corresponding to outflow velocities of approximately $-3{,}800$ km\ s$^{-1}$ and $-10{,}000$ km\ s$^{-1}$, respectively. The primary difference between Model 2 and Model 3 lies in the distribution of the absorber velocities. We also tested a model with a third warm absorber component; however, the additional WA component did not lead to a statistically significant improvement.

To quantitatively assess the relative performance of these models and avoid overfitting, we calculated the Bayesian Information Criterion (BIC; \citealt{Schwarz1978}), defined as:
\[
\mathrm{BIC} = C + k \ln n,
\]
where $C$ is the C-statistic, $k$ is the number of free parameters, and $n$ is the number of spectral bins ($n=137$). The resulting BIC values (Table~\ref{tab:bic_summary}) show a significant improvement with increasing model complexity: Model 1 ($\mathrm{BIC}=584.59$), Model 2 ($\mathrm{BIC}=387.94$), and Model 3 ($\mathrm{BIC}=252.40$). The difference in BIC between successive models was significant. The change in BIC in going from Model 1 to Model 2 was $\sim 200$ and from Model 2 to Model 3 exceeded 80, which clearly favors the more complex model, Model 3 \citep{Kass1995}. 

\begin{table}[h!]
\centering
\caption{Best-fit parameters with symmetric 90\% confidence errors from XSPEC fits to different spectral models for time-average spectral fitting of the entire observation.}
\begin{tabular}{lcc}
\hline
\multicolumn{3}{c}{\texttt{TBabs*(bbody + powerlaw)}} \\
\hline
Component & Parameter (Unit) & Value $\pm$ Error \\
\hline
\texttt{TBabs} & $N_{\mathrm{H}}$ ($\times 10^{20}$ cm$^{-2}$, frozen) & 1.200 \\
\texttt{bbody} & $kT$ (keV) & $0.108 \pm 0.001$ \\
\texttt{bbody} & norm ($\times 10^{-4}$) & $1.656 \pm 0.028$ \\
\texttt{powerlaw} & Photon Index & $1.871 \pm 0.014$ \\
\texttt{powerlaw} & norm ($\times 10^{-3}$) & $5.188 \pm 0.050$ \\
\hline
\multicolumn{2}{l}{C-stat / dof} & 564.91 / 133 \\
\multicolumn{2}{l}{$\chi^2$ / dof} & 247.47 / 133 \\
\hline
\multicolumn{3}{c}{\texttt{TBabs*warmabs*(bbody + powerlaw)}} \\
\hline
\texttt{TBabs} & $N_{\mathrm{H}}$ ($\times 10^{20}$ cm$^{-2}$, frozen) & 1.200 \\
\texttt{warmabs} & Column ($\times 10^{21}$ cm$^{-2}$) & $2.290 \pm 0.191$ \\
\texttt{warmabs} & $\log \xi$ & $1.911 \pm 0.063$ \\
\texttt{warmabs} & $z$ & $-0.065 \pm 0.003$ \\
\texttt{bbody} & $kT$ (keV) & $0.124 \pm 0.002$ \\
\texttt{bbody} & norm ($\times 10^{-4}$) & $1.648 \pm 0.077$ \\
\texttt{powerlaw} & Photon Index & $2.085 \pm 0.019$ \\
\texttt{powerlaw} & norm ($\times 10^{-3}$) & $6.217 \pm 0.107$ \\
\hline
\multicolumn{2}{l}{C-stat / dof} & 353.50 / 130 \\
\multicolumn{2}{l}{$\chi^2$ / dof} & 190.85 / 130 \\
\hline
\multicolumn{3}{c}{\texttt{TBabs*warmabs*warmabs*(bbody + powerlaw)}} \\
\hline
\texttt{TBabs} & $N_{\mathrm{H}}$ ($\times 10^{20}$ cm$^{-2}$, frozen) & 1.200 \\
\texttt{warmabs 1} & Column ($\times 10^{21}$ cm$^{-2}$) & $2.313 \pm 0.294$ \\
\texttt{warmabs 1} & $\log \xi$ & $1.868 \pm 0.073$ \\
\texttt{warmabs 1} & $z$ & $-0.013 \pm 0.003$ \\
\texttt{warmabs 2} & Column ($\times 10^{22}$ cm$^{-2}$) & $1.361 \pm 0.265$ \\
\texttt{warmabs 2} & $\log \xi$ & $3.435 \pm 0.047$ \\
\texttt{warmabs 2} & $z$ & $-0.034 \pm 0.004$ \\
\texttt{bbody} & $kT$ (keV) & $0.135 \pm 0.004$ \\
\texttt{bbody} & norm ($\times 10^{-4}$) & $1.536 \pm 0.083$ \\
\texttt{powerlaw} & Photon Index & $2.148 \pm 0.022$ \\
\texttt{powerlaw} & norm ($\times 10^{-3}$) & $6.841 \pm 0.185$ \\
\hline
\multicolumn{2}{l}{C-stat / dof} & 203.20 / 127 \\
\multicolumn{2}{l}{$\chi^2$ / dof} & 141.05 / 127 \\
\hline
\end{tabular}
\label{tab:time_ave_fit}
\end{table}

\begin{table}[h!]
\centering
\caption{Bayesian Information Criterion (BIC) for time-averaged spectral models. Lower BIC values indicate stronger support for the corresponding model.}
\begin{tabular}{lccc}
\hline
Model & C-stat & Free Params ($k$) & BIC \\
\hline
Model 1 & 564.91 & 4 & 584.59 \\
Model 2 & 353.50 & 7 & 387.94 \\
Model 3 & 203.20 & 10& 252.40 \\
\hline
\end{tabular}
\label{tab:bic_summary}
\end{table}

\subsection{Fitting of Time-resolved Spectra}
All 15 time-resolved spectra were fitted using a best-fit model (\texttt{TBabs*warmabs*warmabs*(bbody + powerlaw)}) to the time-averaged spectrum. However, to improve the stability of the fits and isolate the physical response of the outflowing gas, we fixed the hydrogen column densities of both WA components to the best-fit values obtained from the time-averaged analysis. This assumption is supported by previous high-resolution studies of this source—that the column density remains approximately constant on short time scales \citep[]{kro07, Ogorzalek2022}. Fixing these parameters reduces model degeneracy and allows us to track the ionization state of the absorbers in response to changes in the ionizing continuum.

For each time-resolved spectrum, we allowed the ionization parameters \(\log (\xi)\) of both absorbers and continuum emission component (soft excess black body and power law) parameters to vary freely. Our approach differs from that of \citet{kro07}, who fixed the blackbody temperature in their time-resolved spectral fits. Allowing these components to vary ensures that the fit remains sensitive to both intrinsic continuum variability and spectral shape changes that may accompany variations in the outflow ionization. All time-resolved spectra were fitted independently using XSPEC, and the best-fit parameters for each time bin are listed in Table~\ref{tab:fit_par_table} and visualized in Figure \ref{fig:mainplot}.

\subsection{Cross-correlation and lag analysis}
Using these fitted parameters, obtained from the time-resolved spectral fitting, we then calculated the cross-correlation and possible time lag between the variability of the ionizing source flux and the ionization state of the outflow. 
We note that this time lag has a different physical interpretation from the lag that is frequently used to describe the light travel time between spectral components in  AGN, such as  in reverberation mapping \citep{bla82,kas00,pet04}.

{The cross-correlation coefficient is obtained using the following equation:}
\begin{equation}
C(\tau) = \sum_{t} \left[ \frac{X(t) - \langle X \rangle}{\sigma_X} \right] \left[ \frac{Y(t + \tau) - \langle Y \rangle}{\sigma_Y} \right]
\label{eq:ccf}
\end{equation}
{
Where \( C(\tau) \) is the cross-correlation function at time lag \( \tau \). \( X(t) \) and \( Y(t) \) are the two signals being compared normalized by subtracting their means and dividing by their standard deviations. \( \langle X \rangle \) and \( \langle Y \rangle \) are the mean values of \( X(t) \) and \( Y(t) \), respectively. \( \sigma_X \) and \( \sigma_Y \) are the standard deviations of \( X(t) \) and \( Y(t) \), respectively. \( \tau \) is the time lag  between the ionizing flux and ionization of WA.}

\begin{figure*}
    \centering
    \hbox{
    \includegraphics[width=0.48\linewidth]{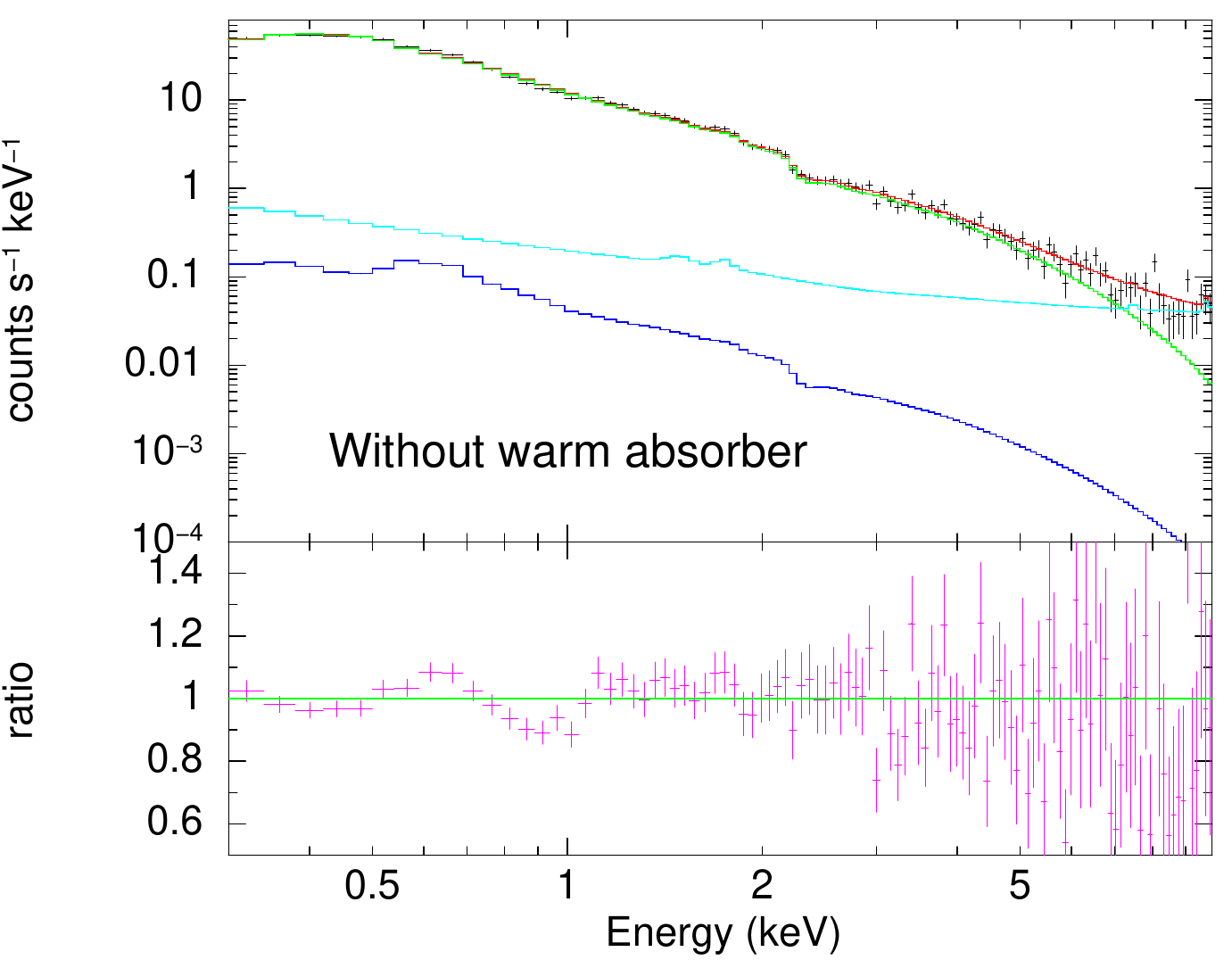}
    \includegraphics[width=0.48\linewidth]{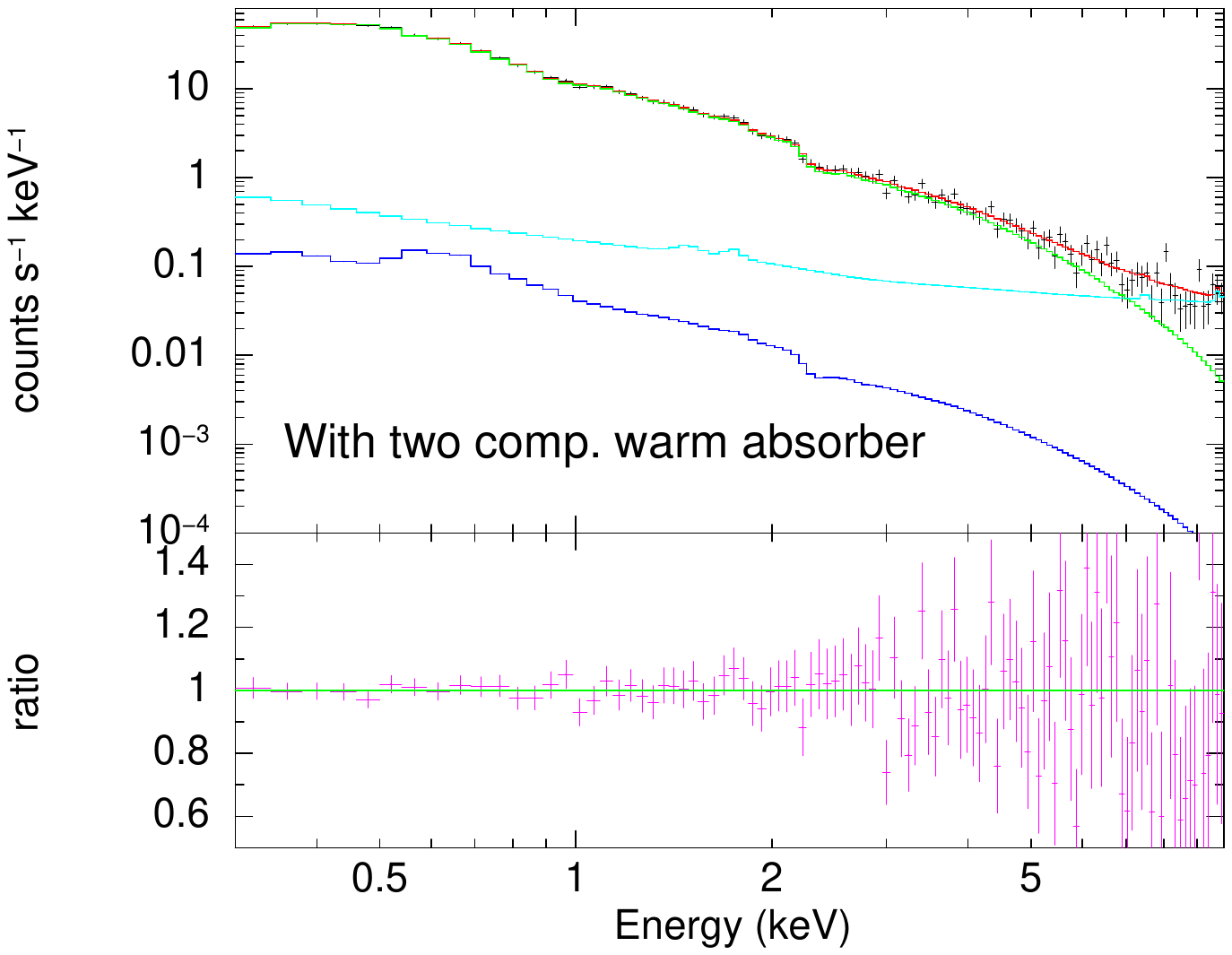}
    }
    \caption{XSPEC fit of a spectrum labeled as s1 in our index, which has an exposure of 993 s. Left panel: fitted with model \textsf{tbabs(bbody+powerlaw)}. Right panel: fitted with a model {\texttt{TBabs*warmabs*warmabs*(bbody + powerlaw)}}. Color code: black points are data, the red curve is the model including background and source model, the green curve is the source model, the cyan-colored curve is the non-sky background, the blue color is the sky background, and the pink points are the ratio plot. The fit was improved significantly with the addition of the two warm absorbers. The $\Delta C  \sim 67$ for this particular fit.}
    \label{fig:spec1_fit}
\end{figure*}

\section{Results and Calculation}
In this section, we present the results obtained from fitting all 15 time-resolved spectra, including their fit parameters, cross-correlation between flux variation and the ionization structure, time lag, and density and location calculation.

\subsection{Fit Parameters for Time-resolved Spectra}

The sensitivity of the NICER spectra to warm absorbers is demonstrated in Figure \ref{fig:spec1_fit}, which compares the fit residuals for models with and without the ionized absorber. The left panel shows a fit using Model 1 (\texttt{TBabs*(bbody+powerlaw)}), while the right panel using Model 3 that includes two WA  (\texttt{TBabs*warmabs*warmabs*(bbody + powerlaw)}). The addition of the warm absorber significantly improves the fit, as seen in the ratio panels below 1 keV. In this particular spectrum (s1), the inclusion of the warm absorber improved the fit by $\Delta C \approx 67$ at the loss of two degrees of freedom.

The best-fit parameters for all 15 time-resolved spectra are summarized in Table \ref{tab:fit_par_table}. The addition of the warm absorber improved the fit for all spectra, with $\Delta C$ values greater than 10 in all cases (only s5 has $\Delta C = 9$), as listed in the last column of the table.

\begin{figure*}
    \centering
    \includegraphics[width=0.96\textwidth]{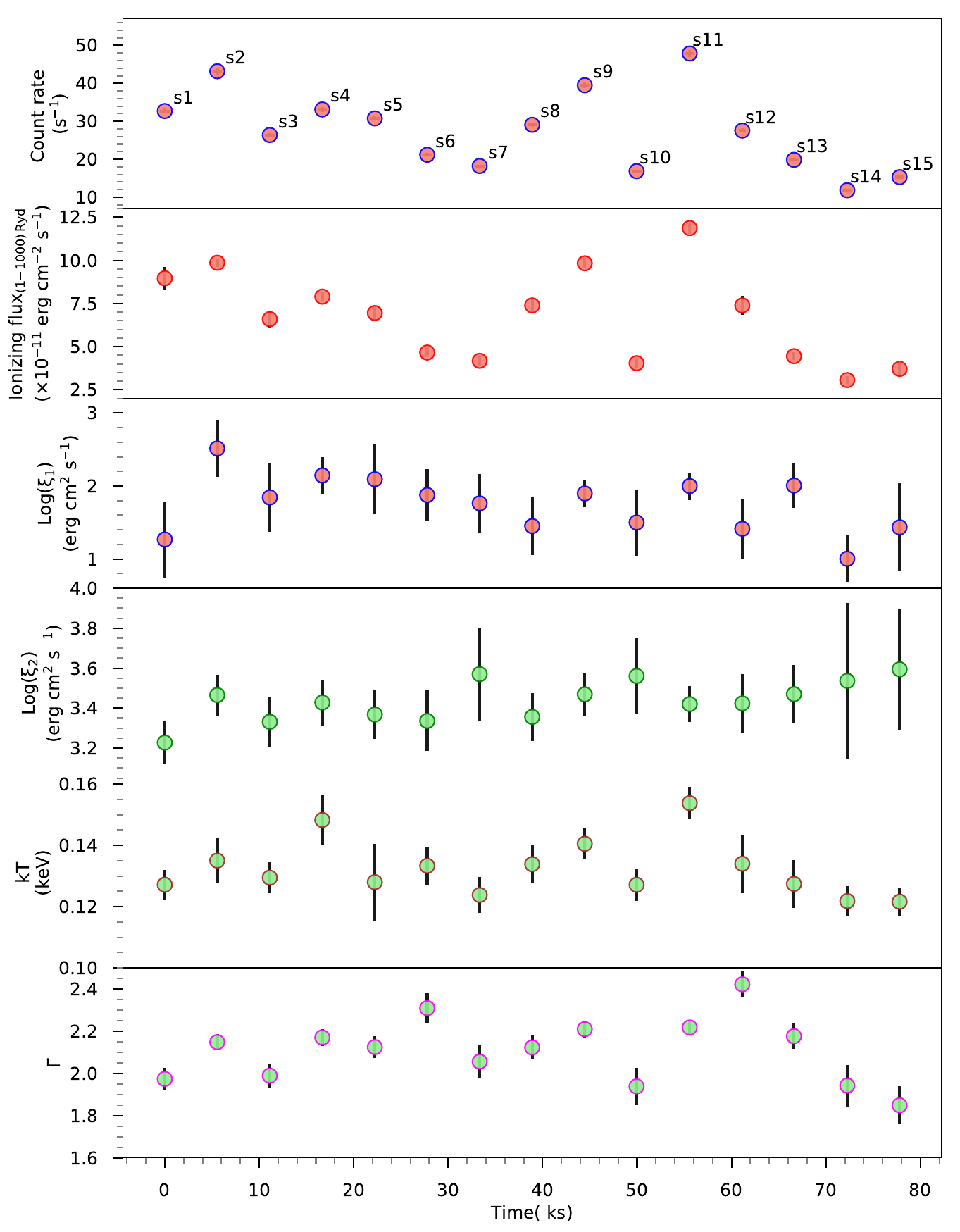}
    \caption{The figure shows the different fitted parameters with their corresponding error bars on the y-axis and time on the x-axis. The exposure time for each spectrum is $\sim 1000$ s. Each data point in the first panel is labeled with numbers s1 to s15 shown in the first panel. $Log(\xi_1)$ and  $Log(\xi_2)$ are the ionization parameters for WA component 1 and component 2, respectively.}
    \label{fig:mainplot}
\end{figure*}

Table \ref{tab:fit_par_table} also presents the continuum parameters for each spectrum: the blackbody temperature $kT$ (keV) and the power-law photon index $\Gamma$, along with their 90\% confidence intervals. The blackbody temperature varies from 0.122 to 0.154 keV, while $\Gamma$ ranges from 1.85 to 2.42 across the 15 spectra. The ionization parameter log($\xi$) of the low-ionization warm absorber component spans from a minimum of 1.00 (s14) to a maximum of 2.51 (s2). For the high-ionization absorber, log($\xi$) ranges from 3.23 (s1) to 3.59 (s15). The unit of $\xi$ is $\mathrm{erg\,cm\,s^{-1}}$. These values and their respective error bars are summarized in Table \ref{tab:fit_par_table} and visualized in Figure \ref{fig:mainplot}. In most cases, the ionization parameters are well constrained within 90\% confidence intervals.

The light curve is shown in the first two panels of Figure \ref{fig:mainplot}, with the first panel representing the count rate and the second panel for the intrinsic ionizing source flux in 1-1000 Ryd. The flux shows significant variability over the observation period, with the flux ranging from $4.04\times 10^{-11}$ erg cm$^{-2}$ s$^{-1}$ in s10 to $11.87\times 10^{-11}$ erg cm$^{-2}$ s$^{-1}$ in  s11. These substantial flux changes occur on timescales of approximately 5500 seconds. The third and fourth panels in Figure \ref{fig:mainplot} display the evolution of the ionization parameters for both warm absorber components across the time-resolved spectra. The bottom two panels illustrate the ionizing continuum parameters, such as blackbody temperature ($kT$) and power-law index ($\Gamma$), both of which exhibit mild variability during the observation.

\begin{table*}[h]
\centering
\caption{Details of the best-fit parameter values for the 15 individual spectra. Each spectrum was fitted using the model 3, \textsf{TBabs$\times${warmabs}$\times${warmabs}(bbody+powerlaw)}.}
\label{tab:fit_par_table}
\begin{tabular}{c c c c c c c c c c}
\hline
seg & Ion. Flux($F$) & log($\xi_1$) & log($\xi_2$) & kT & kT$_{norm}$ & $\Gamma$ & $\Gamma_{norm}$ & C-st./dof & $\Delta C/\Delta dof$ \\
    & $\mathrm{(erg\ cm^{-2}\ s^{-1})}$ & (erg cm s$^{-1}$) & (erg cm s$^{-1}$) & (keV) & $(\times 10^{-4})$ & & ($\times 10^{-3}$) & & (2 WA) \\
\hline
s1 & $8.96 \pm 0.64$ & $1.27 \pm 0.52$ & $3.23 \pm 0.11$ & $0.127 \pm 0.005$ & $3.23 \pm 0.64$ & $1.97 \pm 0.05$ & $8.04 \pm 0.37$ & 91/104 & 67/2 \\
s2 & $9.86 \pm 0.24$ & $2.51 \pm 0.39$ & $3.47 \pm 0.10$ & $0.135 \pm 0.007$ & $1.65 \pm 0.19$ & $2.15 \pm 0.04$ & $10.91 \pm 0.39$ & 129/105 & 49/2 \\
s3 & $6.60 \pm 0.48$ & $1.84 \pm 0.47$ & $3.33 \pm 0.13$ & $0.129 \pm 0.005$ & $1.70 \pm 0.61$ & $1.99 \pm 0.06$ & $6.69 \pm 0.32$ & 99/103 & 44/2 \\
s4 & $7.90 \pm 0.27$ & $2.15 \pm 0.25$ & $3.43 \pm 0.11$ & $0.148 \pm 0.008$ & $1.08 \pm 0.35$ & $2.17 \pm 0.04$ & $8.95 \pm 0.30$ & 107/105 & 17/2 \\
s5 & $6.95 \pm 0.39$ & $2.09 \pm 0.48$ & $3.37 \pm 0.12$ & $0.128 \pm 0.013$ & $1.24 \pm 0.44$ & $2.12 \pm 0.05$ & $8.06 \pm 0.33$ & 111/107 & 9/2 \\
s6 & $4.66 \pm 0.24$ & $1.88 \pm 0.35$ & $3.34 \pm 0.15$ & $0.133 \pm 0.006$ & $1.19 \pm 0.37$ & $2.31 \pm 0.07$ & $4.55 \pm 0.24$ & 115/105 & 20/2 \\
s7 & $4.18 \pm 0.27$ & $1.76 \pm 0.40$ & $3.57 \pm 0.23$ & $0.124 \pm 0.006$ & $1.18 \pm 0.35$ & $2.06 \pm 0.08$ & $4.17 \pm 0.27$ & 108/102 & 25/2 \\
s8 & $7.39 \pm 0.40$ & $1.45 \pm 0.40$ & $3.36 \pm 0.12$ & $0.134 \pm 0.006$ & $2.39 \pm 0.60$ & $2.12 \pm 0.06$ & $7.05 \pm 0.32$ & 108/104 & 28/2 \\
s9 & $9.83 \pm 0.32$ & $1.90 \pm 0.19$ & $3.47 \pm 0.11$ & $0.141 \pm 0.005$ & $1.98 \pm 0.43$ & $2.21 \pm 0.04$ & $10.43 \pm 0.34$ & 82/75 & 22/2 \\
s10 & $4.04 \pm 0.25$ & $1.50 \pm 0.45$ & $3.56 \pm 0.19$ & $0.127 \pm 0.005$ & $1.55 \pm 0.41$ & $1.94 \pm 0.09$ & $3.64 \pm 0.26$ & 100/100 & 23/2 \\
s11 & $11.87 \pm 0.33$ & $2.00 \pm 0.19$ & $3.42 \pm 0.09$ & $0.154 \pm 0.005$ & $2.17 \pm 0.47$ & $2.22 \pm 0.03$ & $12.67 \pm 0.34$ & 114/105 & 35/2 \\
s12 & $7.39 \pm 0.54$ & $1.41 \pm 0.41$ & $3.42 \pm 0.15$ & $0.134 \pm 0.010$ & $2.09 \pm 0.64$ & $2.42 \pm 0.06$ & $6.60 \pm 0.31$ & 132/98 & 21/2 \\
s13 & $4.45 \pm 0.23$ & $2.01 \pm 0.31$ & $3.47 \pm 0.15$ & $0.127 \pm 0.008$ & $0.97 \pm 0.28$ & $2.18 \pm 0.06$ & $5.03 \pm 0.25$ & 121/101 & 17/2 \\
s14 & $3.06 \pm 0.22$ & $1.00 \pm 0.32$ & $3.54 \pm 0.39$ & $0.122 \pm 0.005$ & $1.24 \pm 0.21$ & $1.94 \pm 0.10$ & $2.68 \pm 0.26$ & 80/97 & 20/2 \\
s15 & $3.71 \pm 0.33$ & $1.44 \pm 0.60$ & $3.59 \pm 0.30$ & $0.122 \pm 0.005$ & $1.40 \pm 0.44$ & $1.85 \pm 0.09$ & $3.43 \pm 0.28$ & 81/99 & 11/2 \\
\hline
\end{tabular}
\end{table*}

\subsection{Correlation Between Flux and Ionization 
Parameter}

\begin{figure}
    \centering
    \includegraphics[width=0.45\textwidth]{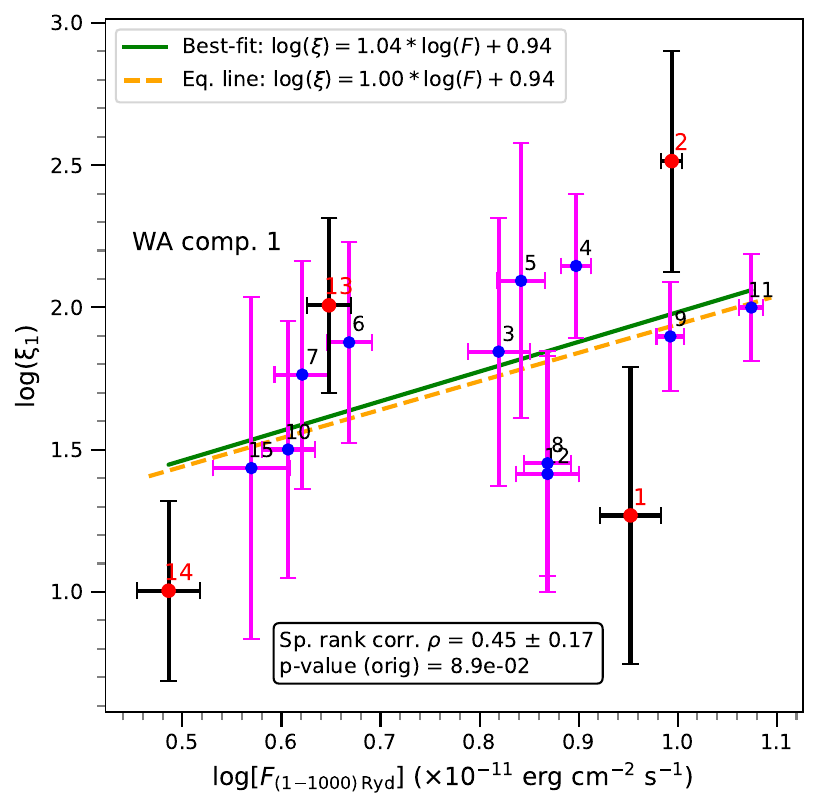}
    \includegraphics[width=0.5\textwidth]{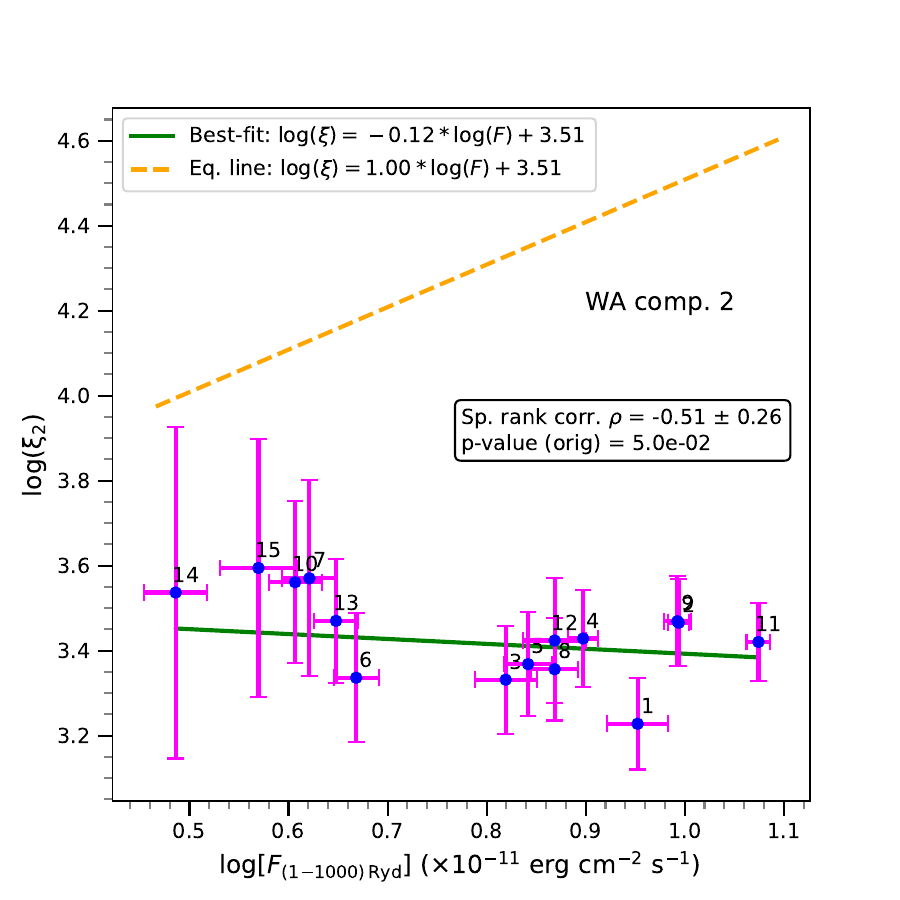}
    \caption{Correlation between the ionization parameter log($\xi$) and the log of the ionizing flux (1--1000 Ryd band) for the two warm absorber components in the time-resolved spectra using Spearman rank correlation method. Top panel: Low-ionization component, showing a moderate positive correlation (Spearman rank coefficient, $\rho=0.45\pm 0.17$) between log($\xi$) and log(flux). The dashed orange line represents the equilibrium condition for photoionized gas, fitted as $\log (\xi) = \log (F) + 0.94$, indicating that this absorber tracks continuum variations and remains close to equilibrium. Bottom panel: High-ionization component, which shows no clear correlation with flux. All measured log($\xi$) values lie below the equilibrium prediction (orange dashed line), suggesting that this component is persistently under-ionized relative to equilibrium.}
    \label{fig:bestfit}
\end{figure}

Figure \ref{fig:bestfit} presents the correlation between the ionization parameter and the ionizing flux ($\textit{F}$) for both warm absorber components across the time-resolved spectra using the Spearman Rank Correlation method. The top panel corresponds to the low-ionization component, while the bottom panel represents the high-ionization component. In both panels, the ionization parameter log($\xi$) is plotted against the log($F$)  in the 1--1000 Rydberg band, calculated from the best-fit spectral model.

The equilibrium condition for photoionized gas is given by $\xi = \frac{4 \pi F}{n}$, which translates to a linear relation in log space: $\log (\xi) = \log (F) + \log (4 \pi / n)$. We include this equilibrium line in our plots (dashed orange line) to directly compare the measured ionization structure with the expectations for photoionization equilibrium. In these plots, the equilibrium line has the form $\log (\xi) = \log (F) + \text{intercept}$, with the y-intercept determined from the best-fited line to all 15 data points.

For the low-ionization absorber (top panel), the data points show a moderate positive correlation between $\log (\xi)$ and $\log (F)$ with the Spearman rank coefficient ($\rho)=0.45\pm 0.17$ and p-value $8.9 \times 10^{-2}$. Most spectra lie close to the equilibrium line, given by $\log (\xi) = \log (F) + 0.94$, indicating that this component responds quickly to variations in the ionizing flux and generally remains in photoionization equilibrium over the observation period. However, s1, s2, s13, and s14 lie above or below the equilibrium lines, indicating the outflow is under- or over-ionized at those times. This result, however, has the different ionization parameters for WA components and resembles the correlation pattern reported by \citet{kro07} for one of the WA components for this source.

In contrast, the high-ionization absorber (bottom panel) shows negative correlation with flux with Spearmanan rank coefficient ($\rho)=-0.51\pm 0.26$ and p-value $8.9 \times 10^{-2}$. All measured ionization parameters for this component consistently lie below the equilibrium prediction, suggesting that the high-ionization phase remains persistently under-ionized relative to the expected equilibrium state. 

\subsection{Cross-Correlation and Time Lag}

To investigate possible time-dependent photoionization signatures, i.e., time lag, we performed cross-correlation analysis between the flux and the ionization parameters at different lags, using equation \ref{eq:ccf}. This cross-correlation calculation is only carried out for low ionization WA components. It is evident that there is no correlation between ionizing flux and the ionization parameters for the high ionization component, as seen in the lower panel of Figure \ref{fig:bestfit}.

\begin{figure*}
    \centering
    \includegraphics[width=1.0\textwidth]{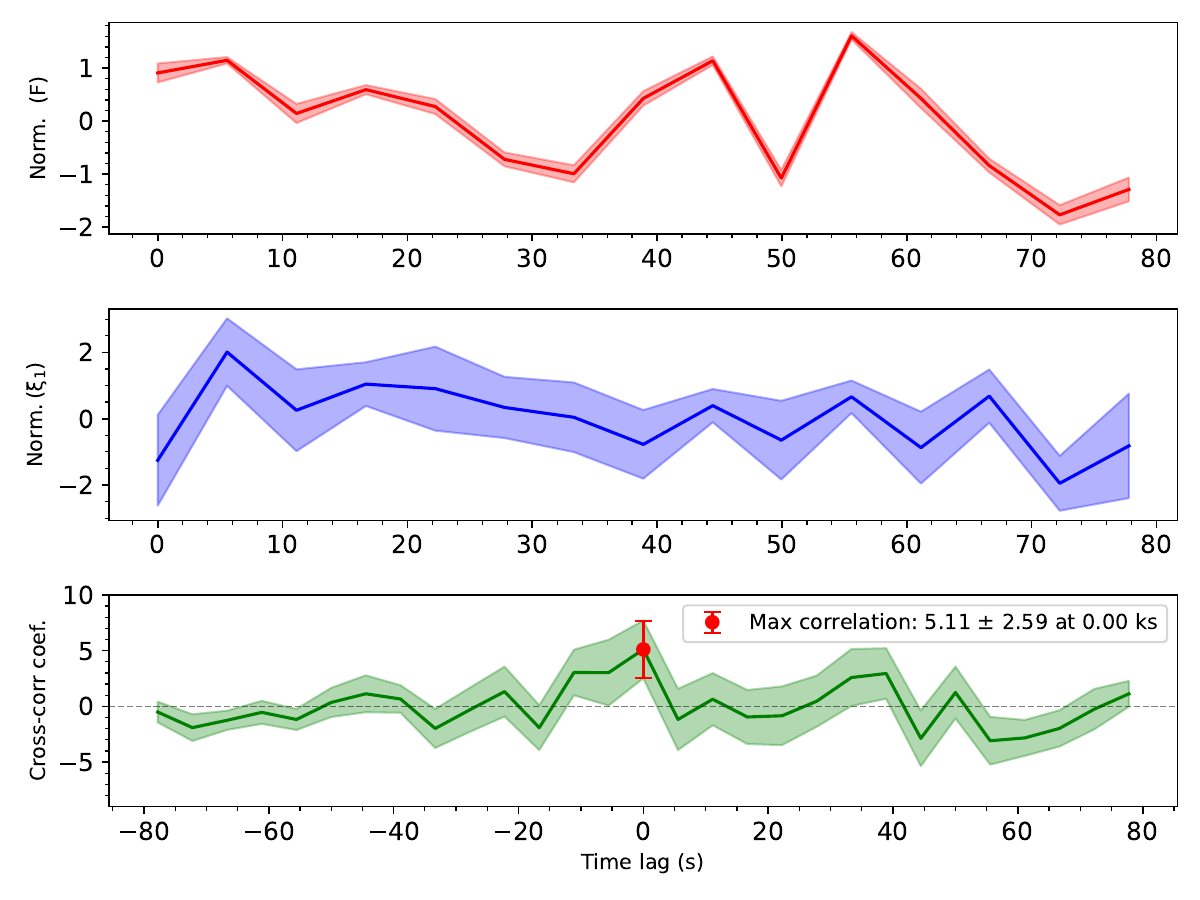}
    \caption{Cross-correlation analysis for the low-ionization warm absorber component. Top panel: normalized light curve with associated error bars. Middle panel: normalized ionization parameter log($\xi_1$) with error bars. Bottom panel: cross-correlation coefficient as a function of time lag between the flux and ionization parameter. The data points are based on 15 spectra, with a time resolution of $\sim 5500$ s, limited by the NICER instrument. Error bars on the cross-correlation coefficient were estimated using Monte Carlo simulations. The maximum correlation coefficient occurs at zero lag, suggesting that the low-ionization warm absorber responds to flux variations and is in photoionization equilibrium.}
    \label{fig:correlation}
\end{figure*}

Figure \ref{fig:correlation} presents the results of the cross-correlation analysis. The top panel shows the flux light curve, the middle panel shows the evolution of the ionization parameter for the low-ionization absorber, and the bottom panel plots the cross-correlation coefficient as a function of time lag. 

The maximum positive correlation is found at zero lag, as shown in the last panel of Figure \ref{fig:correlation}. The absolute value of the  correlation at lag zero is found to be $5.11\pm2.59$. This shows that the ionization parameter responds without measurable delay to changes in the ionizing flux, suggesting the low-ionization warm absorber is largely in photoionization equilibrium throughout the observation. The time resolution of $\sim 5500$ s, set by the NICER sampling, places an upper limit on the response time of the gas: any equilibrium timescale shorter than this cannot be detected.

Given the 90\% confidence interval error bars on the cross-correlation coefficient, the correlation at zero lag is statistically consistent with the highest measured value, and no significant lag is detected. This prevents us from calculating an exact value of gas density. However, this allows us to set up a lower limit of the gas density, owing to the fact that the gas must maintain equilibrium on timescales shorter than 5500 s.

\subsection{Error Estimation in Cross-correlation}
The dark lines in Figure \ref{fig:correlation} represent the measured values, while the shaded regions indicate the associated errors in all three panels. The error bar region is given by the red-colored shadow for the normalized flux as shown in the first panel of Figure \ref{fig:correlation} and the purple-colored shadow for the normalized ionization parameter as shown in the second panel of Figure \ref{fig:correlation}. While the error bars for the flux and ionization parameters were directly derived from the spectral fitting, uncertainties in the cross-correlation function were derived from simulations.

The error bars in the cross-correlation, shown as the green shaded region in the bottom panel of Figure \ref{fig:correlation}, were calculated using Monte Carlo simulations. The cross-correlation itself is determined from the two sets of best-fit values of the quantities being correlated and does not inherently account for the uncertainties in the fitted parameters. To address this, we generated 10,000 simulated light curves and ionization curves by randomly sampling values within the error bars.  For each pair of simulated light and ionization curves, we computed a cross-correlation, resulting in 10,000 correlation curves. The green-shaded region represents the envelope of these 10,000 correlation curves, effectively capturing the uncertainties in the cross-correlation.

\subsection{Time-Evolving Photoionization}
Combining the results shown in Figure \ref{fig:correlation} and the upper panel of Figure \ref{fig:bestfit}, we investigate the response of the warm absorber gas to changes in the ionizing continuum. Since the second warm absorber component does not show a clear response to flux variations, we focus our analysis on the low-ionization component.

The equilibrium timescale $t_{\text{eq}}^{X_i}$ is given by the equation 19 in \citet{sadaula2023}:

\begin{equation}
\label{tim_eq}
t_{\text{eq}}^{X_i} = \frac{1}{n_e \alpha_{\text{rec}}(X_i, T_e) + \gamma_{\text{PI}}}
\end{equation}

where:
\begin{itemize}
    \item $\alpha_{\text{rec}}(X_i, T_e)$ is the radiative recombination coefficient for ion $X_i$ at the electron temperature $T_e$, representing the rate at which the ion recombines with free electrons to form the next lower charge state, $X_{i-1}$.
    \item $n_e$ is the electron density of the gas; higher $n_e$ leads to shorter recombination times and thus a shorter equilibrium timescale.
    \item $\gamma_{\text{PI}}$ is the photoionization rate of ion $X_i$ in the gas.
\end{itemize}

The equilibrium timescale quantifies how long it takes for the ionization state of a specific ion $X_i$ to adjust to changes in the ionizing conditions. In reality, this timescale is influenced by the complex interplay of recombination and photoionization processes across multiple ionic species in the gas. However, a simplified approach considers a single ion that undergoes ionization or recombination depending on the variations in the incident ionizing flux.

This timescale, $t_{\text{eq}}$, is crucial in understanding the behavior of ionized outflows in AGN and warm absorbers, where the gas is exposed to rapidly changing ionizing continua. In low-density environments, the recombination timescale can be longer than the flux variability timescale, leading to persistent departures from ionization equilibrium. Such non-equilibrium conditions can manifest as observable time delays in the ionization state’s response to flux changes or as systematic overionization of the gas compared to equilibrium models.

In this analysis, the maximum correlation at zero lag (Figure \ref{fig:correlation}) suggests that the ionization state of the low-ionization absorber adjusts promptly to changes in the flux, consistent with photoionization equilibrium. The $\sim 5500$ s time resolution of our NICER spectra sets an upper limit on the equilibrium timescale, constraining the gas density to be high enough to allow equilibrium to be reached on timescales shorter than this time resolution we have.

\subsection{Density and Distance Estimates}
The photoionization and recombination timescales are fundamental to determining whether the warm absorber is in equilibrium and, consequently, for estimating the gas density. The photoionization timescale depends on the incident ionizing flux for each ion, while the recombination timescale depends on the electron density and temperature of the gas. In our calculations, we considered only radiative recombination, which typically dominates in the physical conditions of warm absorbers.

The most prominent ionic species contributing to the absorption in the ionized outflow include C VI, N VII, O VII, Ne IX, Mg IX, Si VIII, S IX, Ar XI, and Fe IX. We selected only those ions with ionic fractions greater than 0.3, reflecting the relative abundance of each ion compared to the total number of possible ionization states of that element. The photoionization rates and recombination rate coefficients for these species were computed using equilibrium XSTAR modeling. The input parameters for these calculations, such as ionizing luminosity ($L=1.91\times10^{42}$ erg s$^{-1}$) and $\log(\xi)=1.86$ were taken from the time-averaged spectral fitting and the SED constructed in Section \ref{sec:sed}. The resulting gas temperature was approximately $6.3 \times 10^4$~K. The recombination rate coefficients ($\alpha_{\text{rec}}$) and photoionization rates ($\gamma_{\text{PI}}$) are listed in Table \ref{tab:rec_time}.

\begin{table}[h!]
\caption{Ionic fractions and rates from equilibrium XSTAR modeling for $\log(\xi)=1.86$, and $L=1.91\times10^{42}$ erg s$^{-1}$.}
\label{tab:rec_time}
\centering
\resizebox{\columnwidth}{!}{%
\begin{tabular}{cccccc}
\toprule
Ion & Ionic fr. & RR coef. & Ion. rate \\
    &           & (cm$^3$ s$^{-1}$) & (s$^{-1}$) \\
\midrule
C VI & 0.42 & 8.45E-12 & 1.23E-07 \\
N VII & 0.52 & 1.13E-11 & 4.77E-08 \\
O VII & 0.57 & 1.84E-11 & 8.34E-08 \\
Ne IX & 0.39 & 2.34E-11 & 3.08E-08 \\
Mg IX & 0.30 & 7.91E-11 & 4.39E-07 \\
Si VIII & 0.47 & 1.62E-10 & 1.32E-06 \\
S IX & 0.45 & 1.58E-10 & 8.49E-07 \\
Ar XI & 0.40 & 1.17E-10 & 3.15E-07 \\
Fe IX & 0.34 & 4.92E-10 & 3.66E-06 \\
\bottomrule
\end{tabular}%
}
\end{table}

As summarized in Table \ref{tab:rec_time}, the radiative recombination (RR) coefficient $\alpha_{\text{rec}}$ ranges from $8.45\times10^{-12}$ cm$^3$ s$^{-1}$ for C VI to $4.92\times10^{-10}$ cm$^3$ s$^{-1}$ for Fe IX, reflecting differences in ion charge and temperature. The photoionization rates show similar variation, with Fe IX having the highest rate at $3.66\times10^{-6}$ s$^{-1}$ and N VII among the lowest at $4.77\times10^{-8}$ s$^{-1}$.

From the cross-correlation analysis (Section 5.1), we established that the warm absorber is in equilibrium on timescales of $\lesssim 5500$ s. Using this as the equilibrium timescale $t_{\text{eq}}$ in Equation \ref{tim_eq}, we computed the electron density ($n_e$) for each ion. These values, along with the corresponding absorber distances ($r$) from the central ionizing source, are summarized in Table \ref{tab:ne}.

\begin{table}[h!]
\caption{Electron density $n_e$ and distance $r$ considering each ion, calculated assuming $t_{\mathrm{eq}} = 5500$~s, $L = 1.91 \times 10^{42}$ erg s$^{-1}$, and $\log \xi = 1.86$.}
\label{tab:ne}
\centering
\begin{tabular}{ccc}
\toprule
Ion & $n_e$ (cm$^{-3}$) & $r$ (cm) \\
\midrule
C VI & 2.15E+7 & 3.47E+16 \\
N VII & 1.61E+7 & 4.01E+16 \\
O VII & 9.88E+6 & 5.12E+16 \\
Ne IX & 7.78E+6 & 5.77E+16 \\
Mg IX & 2.29E+6 & 1.06E+17 \\
Si VIII & 1.11E+6 & 1.53E+17 \\
S IX & 1.15E+6 & 1.50E+17 \\
Ar XI & 1.55E+6 & 1.29E+17 \\
Fe IX & 3.62E+5 & 2.67E+17 \\
\bottomrule
\end{tabular}
\end{table}

The calculated electron densities span from $3.62\times10^5$ cm$^{-3}$ for Fe IX to $2.15\times10^7$ cm$^{-3}$ for C VI. These variations reflect the balance between photoionization and recombination rates for each species. Using the relation $\xi = L / (n_e r^2)$, we estimated the absorber distances, which range from $\approx 3.47\times10^{16}$ cm (0.011 pc) for the densest C VI absorber to $\approx 2.67\times10^{17}$ cm (0.087 pc) for Fe IX. The O VII absorber, with $n_e \approx 9.88\times10^6$ cm$^{-3}$, is located at $r \approx 5.12\times10^{16}$ cm (0.017 pc).

\section{Discussion}
The time-resolved analysis of the NICER observation of NGC 4051 provides new insights into the behavior of its warm absorber in response to continuum variability. Despite NICER’s limited spectral resolution, the large effective area and rapid sampling enabled a robust investigation of warm absorber dynamics on short timescales.

\subsection{Time-averaged Spectral Fitting}
Our approach began by fitting the time-averaged spectrum with progressively more complex models to determine the best-fit model, including the warm absorbers. We started with a simple model consisting of neutral absorption, \textsf{TBabs} and continuum emission (\textsf{bbody} + \textsf{powerlaw}), then added one warm absorber component, and finally included a two-component warm absorber model, \textsf{TBabs*warmabs*warmabs*(bbody+powerlaw)}.
All the WA parameters were allowed to vary while we fitted the time-averaged spectrum.

The time-averaged spectral fits revealed that the two-component warm absorber model provided a significantly better fit than the simpler models, as indicated by improvements in the C-statistic and lower BIC values. The best-fit parameters indicate that the two warm absorber components have distinct ionization parameters and column densities, consistent with previous high-resolution studies of NGC 4051 \citep[e.g.,][]{kro07, Ogorzalek2022}. The first, lower-ionization component has $\log(\xi) \sim 1.87$ a column density of $\sim 2.31 \times 10^{21}$~cm$^{-2}$, while the second, higher-ionization component has $\log(\xi) \sim 3.43$ a column density of $\sim 1.36 \times 10^{22}$~cm$^{-2}$. Outflow velocities of both components are consistent with known velocities from grating spectral studies, further supporting the reliability of the two-component WA model, which we have given the name Model 3.

For the time-resolved spectral analysis, we fixed the column densities and outflow velocities of the two warm absorber components to the best-fit values obtained from the time-averaged spectrum. Only the ionization parameters of the warm absorbers were allowed to vary across the time-resolved spectra, tracking the absorber’s response to the variable ionizing continuum. This reduces the number of free parameters in the model, thereby reducing statistical degeneracies in the time-resolved fits.

\subsection{Low-ionization WA Component}
The positive correlation between the ionization parameter of the low-ionization absorber and the ionizing flux (Figure \ref{fig:bestfit}, top panel), combined with the cross-correlation peak at zero lag (Figure \ref{fig:correlation}), suggests that this component remains in photoionization equilibrium throughout the observation on the timescale of $\sim5500$ s. The equilibrium timescale of $\lesssim5500$ s places a lower limit on the gas density of $\gtrsim 10^6$ cm$^{-3}$ (Table \ref{tab:ne}), in line with typical densities for warm absorbers in Seyfert galaxies \citep{kro07, Mizumoto2017, Ogorzalek2022}. The inferred absorber distances, spanning $3\times10^{16}$ cm to $3\times10^{17}$ cm (0.01–0.1 pc), suggest the warm absorber resides within or near the broad-line region, consistent with models of outflows originating in the inner accretion disk \citep{Murray1995, Tombesi2013}.

We note, however, that a few data points—specifically from spectra s1, s2, s13, and s14—lie above or below the equilibrium line, even when 90\% error bars are considered. These deviations likely indicate temporary departures from photoionization equilibrium, possibly due to rapid continuum changes or local density inhomogeneities in the absorber. Despite these, no significant time lag is observed in the cross-correlation analysis (Figure \ref{fig:correlation}), supporting the overall interpretation that the low-ionization gas is largely in equilibrium on the $\sim5500$ s timescale. Consequently, for the purposes of our density and distance estimates, we assume the warm absorber is generally in equilibrium, while acknowledging these out-of-equilibrium fluctuations.

The slight differences in density estimates for different ions (Table \ref{tab:ne}) also highlight the multi-phase structure of the absorber. This variation reflects differences in atomic structure, ionization potential, and photoionization/recombination rates across ions \citep{Kallman2001, Behar2003}. Notably, oxygen (O VII, O VIII) K-shell transitions and L-shell transitions of iron (Fe IX) dominate absorption in warm absorbers and can yield slightly different density estimates because of their distinct recombination and ionization kinetics. Such differences indicate the complex, stratified nature of AGN warm absorbers.

\subsection{High-ionization WA Component}
The high-ionization warm absorber component shows no correlation with the ionizing flux and remains consistently under-ionized relative to equilibrium expectations (Figure \ref{fig:bestfit}, bottom panel). This suggests that this phase is either due to partial shielding or it may be in a collisional plasma, where collisional ionization dominates over photoionization \citep{Ogorzalek2022}. Alternatively, it could be located at larger distances from the central source, where recombination and photoionization processes are decoupled from rapid continuum variability.

\subsection{Limitations}
Although our analysis provides valuable constraints on the warm absorber properties, there are a few caveats that are worth mentioning. First, because of the temporal gaps between NICER pointings ($\sim$4500 s), we lack information about the absorber’s continuous evolution within these intervals. Rapid changes in the ionizing flux could trigger ionization or recombination responses that remain undetected, potentially leading to underestimation of the true variability or misinterpretation of the ionization equilibrium timescale.

The amount of lag we could measure is guided by the time resolution we have available using the cross-correlation technique. Since we have used NICER's data in this work, we have a $\sim 5500$ s of time resolution as mentioned earlier. This means we can measure the time lag of a minimum of $\sim 5500$ s or a multiple of this number. So any lag below $\sim 5500$ s and in between multiples of this number is undetectable.

Our model does not include emission lines from the photoionized gas, focusing solely on the absorption features to minimize the number of free parameters. While this approach isolates the response of the absorber to the ionizing continuum, it does not account for re-emission contributions that could influence the soft X-ray continuum and absorption edges \citep{Mehdipour2011,Mizumoto2017}. Incorporating emission lines would provide a more complete picture of the reprocessing in the warm absorber but requires higher spectral resolution data to avoid degeneracies in the fit.

\subsection{Implications for Outflows}
These findings underscore the utility of time-resolved X-ray spectroscopy, even with moderate spectral resolution, for constraining the density and location of warm absorbers. The warm absorber in NGC 4051 appears to originate within or near the BLR, consistent with disk-wind or stratified outflow models. Our density estimates for the low-ionization phase are based on variability measured over a single day, corresponding to the duration of our NICER observation. While these snapshot measurements provide robust lower limits on the gas density and upper limits on the distance, they represent only a short-term view of the warm absorber’s properties. Over longer periods from months to years, the warm absorber parameters can vary significantly, reflecting the evolving physical conditions in the outflow. Monitoring campaigns with higher spectral resolution and multi-epoch observations have shown that both the density and location of warm absorbers can evolve in response to changes in the accretion disk luminosity and structure. 

The low-ionization phase in our data maintains photoionization equilibrium with the variable continuum, while the high-ionization phase may represent a distinct collisional plasma or a slower-responding, lower-density component. Such multiphase outflows are promising candidates for contributing to AGN feedback, potentially influencing the evolution of the host galaxy’s interstellar medium. Future observations with higher cadence and multiwavelength coverage will be crucial to disentangle these temporal effects and track the long-term evolution of warm absorber properties in AGN.
\vspace{1cm}
\section{Conclusion}
This study advances our understanding of the density, location, and time-dependent behavior of warm absorbers in AGN, using time-resolved X-ray spectroscopy as a powerful tool for investigating the properties of ionized outflows.

Using NICER observations of NGC 4051, we conducted a time-resolved analysis of the warm absorber’s response to continuum variability, tracking absorber variability on $\sim 5500$ s timescales. In these time-resolved fits, the column density and outflow velocity were fixed at their best-fit values from the time-averaged spectrum, allowing us to robustly measure the evolution of the ionization parameter (log~$\xi$) for the low-ionization warm absorber. We find that log~$\xi$ varies from 1.00 to 2.51 across the 15 time-resolved spectra, generally tracking changes in the ionizing flux. The strong positive correlation between log~$\xi$ and flux—combined with the absence of a measurable lag—indicates that the low-ionization component remains in photoionization equilibrium on these timescales. This constrains the equilibrium timescale to $\lesssim 5500$ s and implies a lower limit on the electron density of $\gtrsim 9 \times 10^6$ cm$^{-3}$, consistent with warm absorber densities found in other Seyfert galaxies. The inferred distance of the warm absorber from the ionizing source, $\lesssim 7.02 \times 10^{16}$ cm ($\sim 0.023$ pc), places it within or near the broad-line region, supporting models in which warm absorbers are launched from the inner accretion disk.

In contrast, the high-ionization component shows no clear correlation with flux, remaining persistently under-ionized relative to equilibrium predictions. This may indicate a longer response timescale, lower density, or a different physical origin—such as collisional ionization in a hotter plasma phase.

These findings highlight the value of time-resolved X-ray spectroscopy, even with moderate spectral resolution, for constraining the density and location of warm absorbers. By carefully modeling the variability of the ionization parameter in response to continuum changes, we can disentangle key physical properties of the ionized outflows. Future observations with higher spectral resolution and improved time sampling will be critical for resolving the detailed structure and dynamics of AGN warm absorbers and understanding their role in AGN feedback and galaxy evolution.

\section{Future Prospect}
Understanding how supermassive black holes (SMBHs) interact with their host galaxies  \citep{gebhardt2000,ferrarese2000} remains a key question in astrophysics, with warm ionized absorbers considered critical agents in this feedback process. However, fundamental properties such as density and location remain challenging to determine independently due to degeneracies between them. Time-resolved spectroscopy can help break these degeneracies, but current instruments like XMM-Newton and $Chandra$ require long exposures for sufficient signal-to-noise, limiting their time resolution. In this study, we demonstrated that NICER’s high-cadence observations allow us to put a constraint on warm absorber density by analyzing time-resolved spectral fits and their correlation with continuum flux variability, even within the limitations imposed by Earth occultations. The future X-ray missions with improved sensitivity and uninterrupted observing capabilities, such as the New Advanced Telescope for High-Energy Astrophysics (NewAthena), which has an effective area approximately an order of magnitude larger than XMM Newton, will be able to do time-resolved spectroscopy studies of WA of around 100 s.  The new proposed NASA probe mission, Advanced X-ray Imaging Satellite (AXIS) telescope, will have an effective area approximately four times that of XMM-Newton. This allows us to do the time-resolved spectroscopic studies of ionized absorbers in the order of a few hundred seconds.  This enables more precise constraints on warm absorber properties and a deeper understanding of AGN outflows.

\subsection*{Acknowledgments}

We are grateful to the anonymous referee for their constructive comments and suggestions, which helped improve the clarity and robustness of this manuscript.

\bibliographystyle{aasjournal}
\bibliography{tdp}

\end{document}